\documentclass[superscriptaddress, twocolumn]{revtex4-1}

\usepackage{graphicx}
\usepackage{epstopdf}
\usepackage{amsmath}
\usepackage{color}
\usepackage{float}
\usepackage{textgreek}
\usepackage{lineno}

\begin{document}
\title{Self-corrected chip-based dual-comb spectrometer}

\author{Nicolas Bourbeau H\'ebert}  \email{nicolas.bourbeau-hebert.1@ulaval.ca}
\affiliation{Centre d'optique, photonique et laser, Universit\'{e} Laval, Qu\'{e}bec, QC, G1V 0A6, Canada}
\author{J\'er\^ome Genest}
\affiliation{Centre d'optique, photonique et laser, Universit\'{e} Laval, Qu\'{e}bec, QC, G1V 0A6, Canada}
\author{Jean-Daniel Desch\^enes}
\affiliation{Centre d'optique, photonique et laser, Universit\'{e} Laval, Qu\'{e}bec, QC, G1V 0A6, Canada}
\author{Hugo Bergeron}
\affiliation{Centre d'optique, photonique et laser, Universit\'{e} Laval, Qu\'{e}bec, QC, G1V 0A6, Canada}
\author{George Y. Chen}
\affiliation{Laser Physics and Photonics Devices Laboratory, Future Industries Institute, and School of Engineering, University of South Australia, Mawson Lakes, SA 5095, Australia}
\author{Champak Khurmi}
\affiliation{Laser Physics and Photonics Devices Laboratory, Future Industries Institute, and School of Engineering, University of South Australia, Mawson Lakes, SA 5095, Australia}
\author{David G. Lancaster}
\affiliation{Laser Physics and Photonics Devices Laboratory, Future Industries Institute, and School of Engineering, University of South Australia, Mawson Lakes, SA 5095, Australia}

\date{\today}


\begin{abstract}
We present a dual-comb spectrometer based on two passively mode-locked waveguide lasers integrated in a single Er-doped ZBLAN chip. This original design yields two free-running frequency combs having a high level of mutual stability. We developed in parallel a self-correction algorithm that compensates residual relative fluctuations and yields mode-resolved spectra without the help of any reference laser or control system. Fluctuations are extracted directly from the interferograms using the concept of ambiguity function, which leads to a significant simplification of the instrument that will greatly ease its widespread adoption. Comparison with a correction algorithm relying on a single-frequency laser indicates discrepancies of only 50 attoseconds on optical timings. The capacities of this instrument are finally demonstrated with the acquisition of a high-resolution molecular spectrum covering 20 nm. This new chip-based multi-laser platform is ideal for the development of high-repetition-rate, compact and fieldable comb spectrometers in the near- and mid-infrared.
\end{abstract}

\maketitle

\section{Introduction}

The development of advanced spectrometers leads to new insights into science \cite{cossel2012broadband, li2008laser, truong2015accurate} and enables improvements in production environments through industrial process control \cite{berntsson2002quantitative, funke2003techniques}. Spectrometer development took a substantial step forward with the emergence of frequency combs; their broad and regularly spaced modal structure makes them excellent sources to achieve active spectroscopy with unrivalled frequency precision \cite{maslowski2016surpassing, stowe2008direct, foltynowicz2013cavity, marian2005direct, holzwarth2000optical}. However, this precision can only be captured if the comb modes are spectrally resolved. 

Dual-comb spectroscopy \cite{coddington2016dual} is one of the few techniques able to resolve a complete set of dense comb modes. It maps the optical information to the more accessible radio-frequency (RF) domain using mutually coherent combs having slightly detuned repetition rates. Their coherence is typically ensured by phase locking both combs together or to external references \cite{coddington2008coherent, baumann2011spectroscopy}, by using a post-correction algorithm based on auxiliary lasers \cite{deschenes2010optical, roy2012continuous}, or by using an adaptive sampling scheme \cite{ideguchi2014adaptive}. However, all these approaches rely on external signals and additional hardware, which adds a significant layer of complexity to the dual-comb instrument.

Some laser designs have recently been proposed to generate two slightly detuned combs from the same cavity in order to force a certain level of mutual coherence enabled by the rejection of common-mode noise. Most are based on non-reciprocal cavities that induce a repetition rate difference \cite{ideguchi2016kerr, mehravar2016real, gong2014polarization}. The generation of two combs with different central wavelengths was also reported \cite{zhao2016picometer, chang2015dual}, but this avenue requires an additional step to broaden the lasers and obtain enough spectral overlap. However, having two pulse trains sharing the same gain and mode-locking media, which are both highly nonlinear, is worrisome as it could introduce delay-dependant distortions in interferograms (IGMs). Indeed, a pair of pulses overlapped in a nonlinear medium could be significantly different from another pair interacting separately with the medium. In fact, the long-known colliding-pulse laser \cite{fork1981generation} exploits this effect to shorten the duration of its pulses. Dual-comb generation using two cavities integrated on a single platform avoids this concern and has been demonstrated with few-mode semiconductor combs \cite{link2015dual, link2016dual, villares2015chip, rosch2016chip}.

Even those common-mode designs have difficulty to yield combs with sufficient relative stability to allow coherent averaging of data \cite{ideguchi2016kerr, mehravar2016real, zhao2016picometer}. Thus, additional hardware and signals are still needed to track and compensate residual drifts. An interesting idea was recently suggested to extract those drifts directly from the IGMs using predictive filtering \cite{burghoff2016computational}. Since it comes down to tracking the time-domain signal using a model made from the sum of all comb modes, the effectiveness of this approach still has to be demonstrated for cases with several thousand modes and where signal is only available momentarily in bursts near zero path difference (ZPD).

In this paper, we present a standalone and free-running dual-comb spectrometer based on two passively mode-locked waveguide lasers (WGLs) \cite{champak, schlager2003passively, beecher2010320, thoen2000erbium} integrated in a single glass chip. This mutually stable system allows to fully resolve the comb modes after using a new algorithm that corrects residual relative fluctuations estimated directly from the IGMs. Thus, no single-frequency lasers, external signals or control electronics are required to retrieve the mutual coherence, which tremendously simplifies the instrument. The design we use for this demonstration is also original and consists of two ultrafast-laser-inscribed waveguides \cite{gross2013femtosecond, minoshima2001photonic, davis1996writing, choudhury2014ultrafast} in a chip of Er-doped ZBLAN, forming two mechanically coupled, but optically independent, laser cavities. Lasers are mode-locked using two distant areas of the same saturable absorber mirror (SAM). This design avoids nonlinear coupling between combs while maximizing their mutual stability. We use the instrument to collect a 20-nm-wide absorption spectrum of the $2\nu_{3}$ band of hydrogen cyanide (H\textsuperscript{13}C\textsuperscript{14}N). The high quality of the spectral data (acquired in 71 ms) is validated by fitting Voigt lines that return parameters in close agreement with published values.


\section{Instrument design}
\label{sec:instru}
WGLs are remarkably well adapted to support dual-comb spectrometers. Indeed, several waveguides are typically available on a chip, they offer a much lower cavity dispersion than fibre lasers, thanks to the short propagation through glass, which facilitates mode-locking, and their small size is compatible with the market's demand for small-footprint instruments. Furthermore, the transparency of ZBLAN from visible to mid-infrared allows for a broad range of emission wavelengths to be supported \cite{smart1991cw, palmer2013high, lancasterer3+, lancaster2015holmium, lancaster2013efficient}. Finally, rare-earth-doped glasses have proven to be excellent candidates for the generation of low-noise frequency combs of metrological quality. WGLs thus seem to be an obvious choice for the centrepiece of a dual-comb platform.

\begin{figure}[htbp]
\centering
\includegraphics[width=\linewidth]{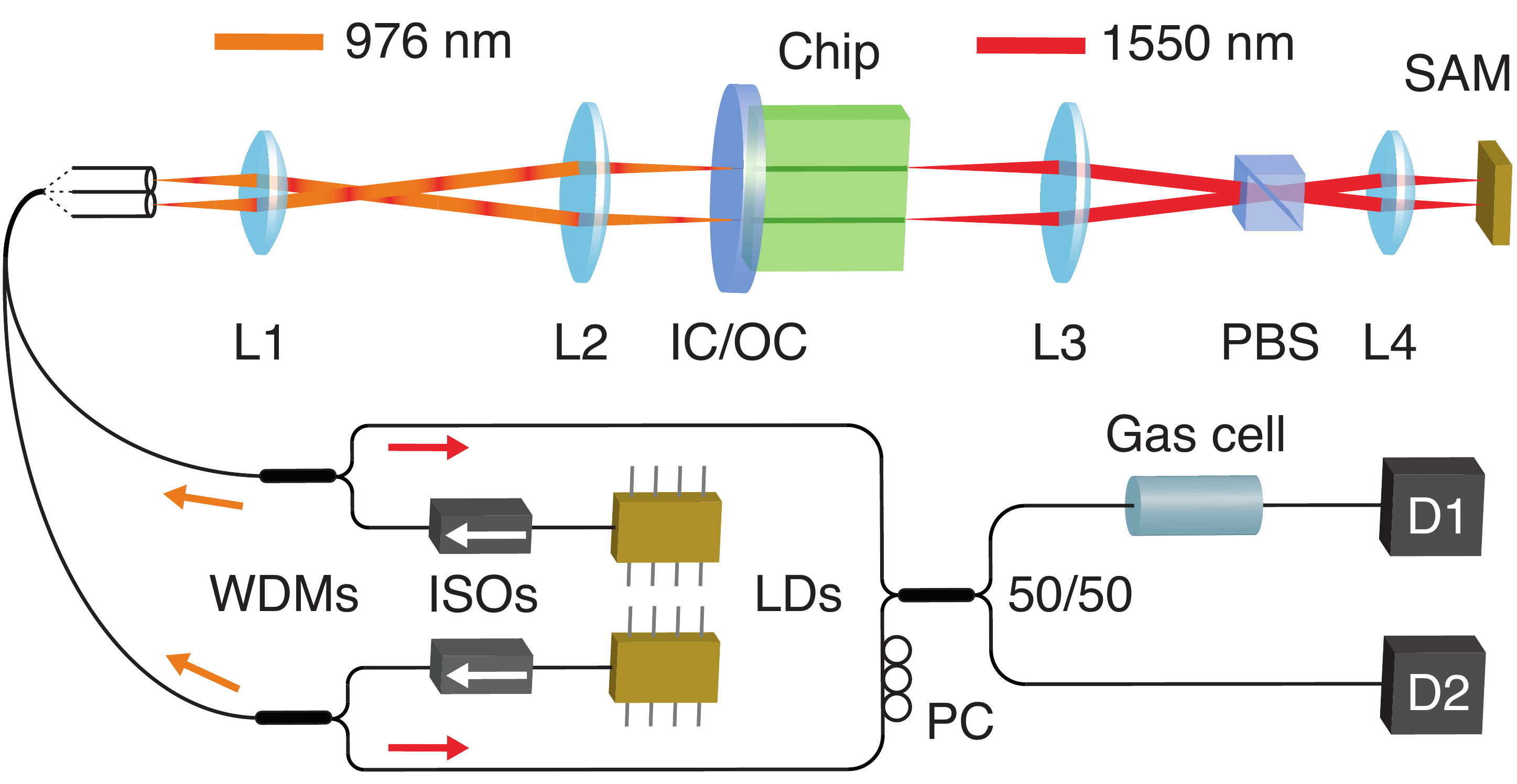}
\caption{Schematic of the dual-comb spectrometer. Two parallel waveguides laser-inscribed in an Er-doped ZBLAN glass chip are used to generate a pair of frequency comb lasers, which are mode-locked with a common SAM. The fibre-coupled outputs are used to perform dual-comb spectroscopy. PC: polarization controller, D1 and D2: detectors.}
\label{fig:setup}
\end{figure}

Figure \ref{fig:setup} shows a schematic of the dual-comb spectrometer, whose design is inspired by the single-cavity mode-locked WGL presented in \cite{champak}. It revolves around a 13-mm-long ZBLAN glass chip containing several laser-inscribed waveguides \cite{gross2013femtosecond} with diameters ranging from 30 to 55 \textmu m, which all support single-transverse-mode operation. The glass is doped with 0.5 mol\% Er\textsuperscript{3+}, acting as the active ion, 2 mol\% Yb\textsuperscript{3+}, which enhances pump absorption \cite{nagamatsu2004influence}, and 5 mol\% Ce\textsuperscript{3+}, which reduces excited-state absorption in Er\textsuperscript{3+} \cite{nagamatsu2004influence}.

Two laser diodes (LDs) (Thorlabs BL976-PAG900), each capable of producing around 900 mW of single-transverse-mode power at 976 nm, are used to pump the chip. They go through separate isolators (ISOs) (Lightcomm HPMIIT-976-0-622-C-1) and the end of the output fibres are stripped, brought in contact along their side, and sandwiched between two glass slides with glue. The fibres are therefore held in place with a distance of 125 \textmu m between cores and with the end facets lying in the same plane, which is just sticking out of the sandwich. 

The output plane is imaged onto the chip with a pair of lenses (L1 and L2) arranged in an afocal configuration to couple the pump beams into a pair of waveguides separated by 600 \textmu m (centre-centre). The lenses are chosen so that the ratio of the focal lengths best matches the required magnification set by the distance between waveguides and that between fibres (4.8 in this case). A software-assisted optimization of distances between components is performed for the chosen lenses in order to maximize coupling. Two parallel waveguides having diameters of 45 and 50 \textmu m are selected since we observe that they yield the best efficiencies as a result of a good balance between mode matching and pump confinement. The waveguides' large area ensures a low in-glass intensity, which increases the threshold for undesirable nonlinear effects \cite{champak}.

An input coupler (IC), which also acts as an output coupler (OC), is butted against the left side of the chip in order to let the pump in ($T_{976} > 95\%$) and to partially reflect the signal ($R_{1550} = 95\%$). On the other side, a pair of anti-reflective coated lenses (L3 and L4) arranged in an afocal configuration is used to image the waveguide modes onto a SAM (Batop SAM-1550-15-12ps) with a magnification of 0.16. This size reduction increases the fluence on the SAM, and thus its saturation, which permits the passive mode-locking of the lasers. A polarization beam splitter (PBS) is placed between lenses L3 and L4 to allow a single linear polarization. Both cavities make use of the same components, which ensures maximum mutual stability.

The resulting mode-locked frequency combs exit their respective cavity at the OC and travel back towards the fibres to be collected. They are separated from the counter-propagating pumps with wavelength-division multiplexers (WDMs) (Lightcomm HYB-B-S-9815-0-001), which also include a stage of isolation for the signal wavelength. This conveniently gives two fibre-coupled frequency comb outputs that can be mixed in a 50/50 fibre coupler to perform dual-comb spectroscopy.  Each cavity generates  $\sim 2$ mW of comb power, of which around $10\%$ is successfully coupled in the fibres. This is due to the alignment being optimized for the pump wavelength, thus the efficiency could be improved with an achromatic imaging system. Nevertheless, this level of power is more than sufficient for laboratory-based spectroscopy.

\begin{figure}[htbp]
\centering
\includegraphics[width=\linewidth]{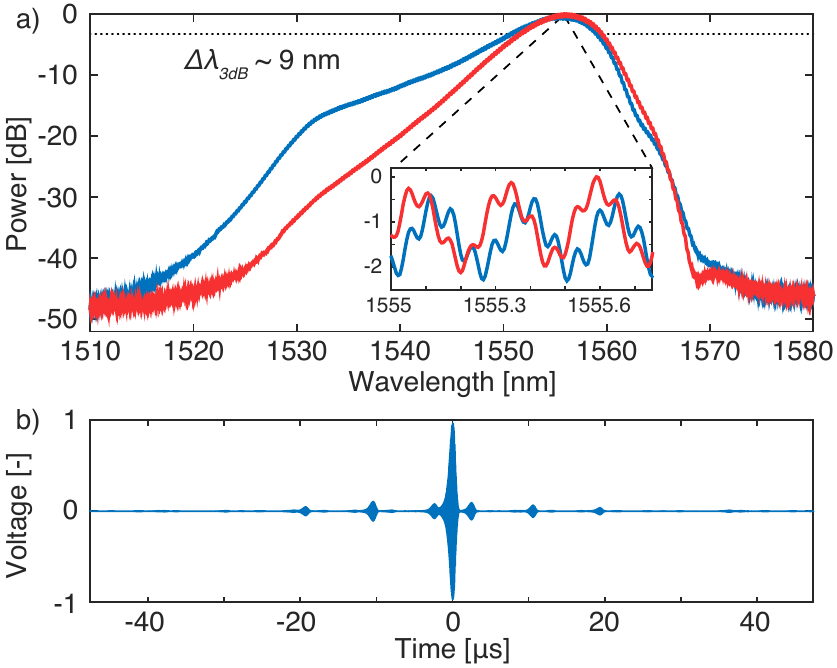}
\caption{a) Spectrum of each frequency comb. They span $9$ nm around 1555 nm and show excellent spectral overlap. The resolution of the optical spectrum analyzer is set to 0.2 nm (0.03 nm for the inset). b) Averaged IGM obtained with a self-corrected sequence of IGMs.}
\label{fig:OSA}
\end{figure}

Figure \ref{fig:OSA}a) shows the power spectrum of each comb, as measured with an optical spectrum analyzer (Anritsu MS9470A). Their 3 dB bandwidth ($\Delta \lambda_{3dB}$) spans approximately 9 nm around 1555 nm and they show excellent spectral overlap. A zoomed view reveals spectral modulation that is identified as parasitic reflections taking place on the left surface of the OC and on the right surface of the chip. Even though anti-reflective coatings are deposited on those surfaces, the weak echoes are re-amplified through the chip and come out with non-negligible power. This issue can be solved with an angled chip \cite{lancaster2015holmium} and a wedged OC.

The repetition rate ($f_r$) of each comb is $822.4$ MHz and their repetition rate difference ($\Delta \! f_r$) is $10.5$ kHz. This yields a beat spectrum fully contained within a single comb alias. Its central frequency is adjustable by varying one of the pump diodes' power. As for $\Delta \! f_r$, it is mostly determined by the slight optical path differences through lenses and, potentially, through waveguides. Indeed, their diameters differ and this affects their effective refractive indices. Tuning $\Delta \! f_r$ is possible by slightly adjusting the alignment of optical components. Figure \ref{fig:OSA}b) shows an averaged IGM obtained with a sequence of IGMs self-corrected using the algorithm presented in the next section. Small pulses on either side of the ZPD burst correspond to the parasitic reflections mentioned earlier.

\begin{figure}[htbp]
\centering
\includegraphics[width=\linewidth]{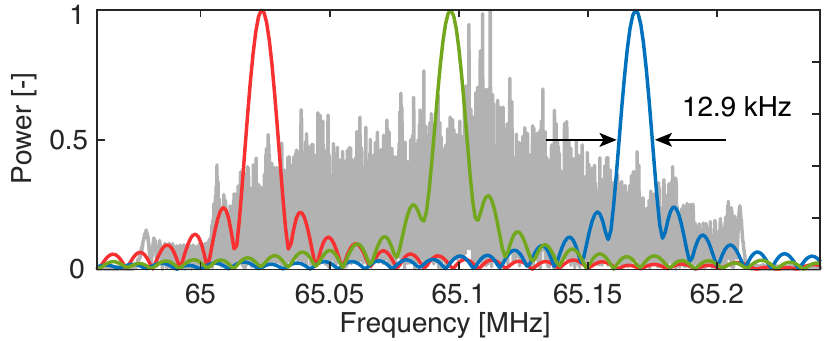}
\caption{Beat note between two comb modes, one from each comb, for different observation times: 71 ms (grey) and $1/\Delta \! f_r \sim 95$ \textmu s (red, green, blue). Coloured traces are computed from three different sections of the 71 ms measurement. Their width is close to the transform limit, which indicates that the dual-comb platform is mutually stable on a $1/\Delta \! f_r$ timescale.}
\label{fig:moderaw}
\end{figure}

The mutual stability of the dual-comb platform is evaluated using the beat note between two comb modes, one from each comb, measured through an intermediate continuous-wave (CW) laser. Figure \ref{fig:moderaw} shows the beat note computed from a 71 ms measurement (grey), which corresponds to the digitizer's memory depth at 1 GS/s, along with beat notes computed from three different sections of duration $1/\Delta \! f_r \sim 95$ \textmu s belonging to the longer measurement (red, green, blue). Coloured traces are nearly transform-limited since their width ($\sim 12.9$ kHz) approaches the bandwidth of a rectangular window ($1.2 \Delta \! f_r = 12.6$ kHz). This means that the dual-comb platform is stable to better than $\Delta \! f_r$ on a $1/\Delta \! f_r$ timescale, which consists of a key enabler for the self-correction algorithm presented below. However, the beat note's central frequency oscillates on a slower timescale and turns into the wider grey trace ($>10\Delta \! f_r$) after 71 ms integration. This is mostly due to vibrations that slightly change the coupling of the pumps into the waveguides as well as the alignment of intra-cavity components.

\section{Self-correction}
\label{sec:algo}

Although nothing forces the combs to settle individually at specific frequencies, the platform presented here is designed to provide them with mutual stability. Therefore, the frequency difference between pairs of comb modes is more stable than their absolute frequencies. This is exactly what is required for mode-resolved dual-comb spectroscopy since the measured beat spectrum is a new RF comb with modes sitting at those differential frequencies. In order to reach a specified spectral resolution, the stability constraints on the RF comb need to be more severe than those on the optical combs by a factor equal to the compression ratio between the optical and RF domains ($f_r / \Delta \! f_r$).

We can define the RF comb with only two parameters: its spectral offset and its spectral spacing. Mathematically, the RF modes are found at frequencies $f_n = f_c + n \Delta \! f_r$, where $f_c$ is the frequency of the mode closest to the carrier frequency (the spectrum's centre of mass) and $n$ is the mode index. It is judicious to define the comb around $f_c$ since this reduces the extent of $n$, which acts as a lever on $\Delta \! f_r$, and thus increases the tolerance on the knowledge of this parameter.  Of course, $f_n$ is a time-dependent quantity since residual fluctuations $\delta \! f_{c}(t)$ and $\delta \! \Delta \! f_r(t)$ remain despite our instrument design. The modes' frequencies can thus be described at all times with

\begin{equation}
f_n(t) =  [f_{c} + \delta \! f_{c}(t)] + n [\Delta \! f_r + \delta \! \Delta \! f_r(t)]
\label{eq:electriccalcomb}
\end{equation}

When measuring dual-comb IGMs generated with free-running combs, it is required that we estimate and compensate those fluctuations. This allows reaching the spectral resolution made available by the optical combs and it opens the door to coherent averaging \cite{coddington2016dual} by yielding mode-resolved spectra. We show here that it is possible to extract the residual fluctuations directly from the IGMs by making use of the cross-ambiguity function \cite{woodward2014probability}, a tool initially developed for radar applications. This tool is closely related to the cross-correlation, but besides revealing the time delay $\tau$ between two similar waveforms, it also reveals their frequency offset $f_o$. More specifically, the cross-ambiguity function gives a measure of the similarity of two waveforms, $A_1(t)$ and $A_2(t)$, as a function of $\tau$ and $f_o$. It is given by

\begin{equation}
\chi_{1,2}(\tau, f_o ) = \int_{-\infty}^{\infty} A_1(t) \, A_2^*(t+\tau) \, \exp(2\pi i f_o t)dt
\label{eq:ambiguity}
\end{equation}

\noindent where $^*$ denotes complex conjugation. In the presence of chirp, an uncompensated frequency shift can affect the apparent delay between waveforms as retrieved by the more familiar cross-correlation method. Hence, it is important to retrieve those two parameters simultaneously from the point of maximum similarity on an ambiguity map, that is where $|\chi_{1,2}(\tau, f_o)|$ is maximum.

\begin{figure}[htbp]
\centering
\includegraphics[width=\linewidth]{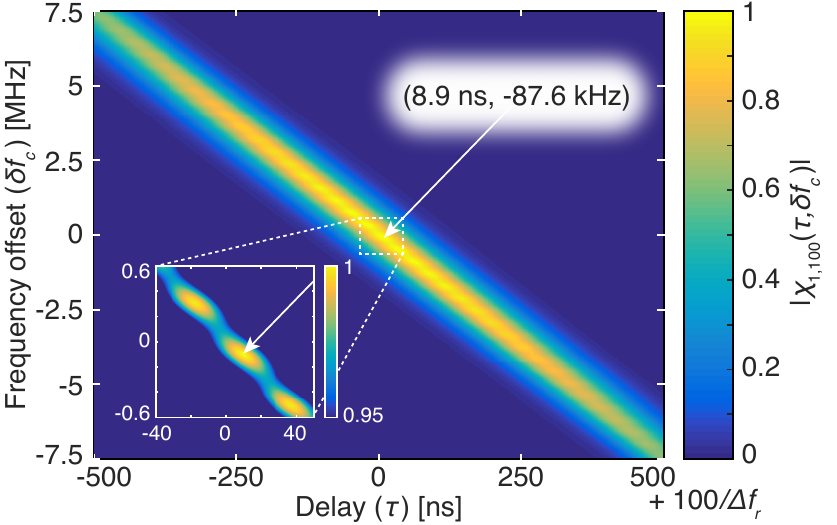}
\caption{Normalized ambiguity map generated from measured IGMs for the case $k=100$. The delay axis is centered on the expected delay $k/\Delta \! f_r$. The coordinates $(\tau_k, \delta \! f_{c,k})$ at the point of maximum similarity are also given. Notice how the function takes large values along an oblique line, which illustrates the coupling between apparent delay and frequency offset when working with chirped IGMs.}
\label{fig:ambiguity}
\end{figure}

For a given dual-comb IGM stream, we can compute $\chi_{1,k}(\tau,\delta \! f_c )$ between the first and $k^{th}$ ZPD bursts, where $f_o$ takes the form of a frequency offset $\delta \! f_c$ relative to the first burst's $f_c$ in that specific context. The values $\tau_k$ and $\delta \! f_{c,k}$ at the maximum of each calculated ambiguity map reveal the instantaneous fluctuations sampled at each ZPD time. Indeed, time delays $\tau_k$ translate into fluctuations $\delta \! \Delta \! f_r(t)$, while $\delta \! f_{c,k}$ are samples from $\delta \! f_c(t)$. Figure \ref{fig:ambiguity} shows an ambiguity map generated from measured IGMs for the case $k=100$. Only ZPD bursts and their slightly overlapping adjacent echoes are used for the calculation. The latter are responsible for the weak modulation on the ambiguity map.

\begin{figure}[htbp]
\centering
\includegraphics[width=\linewidth]{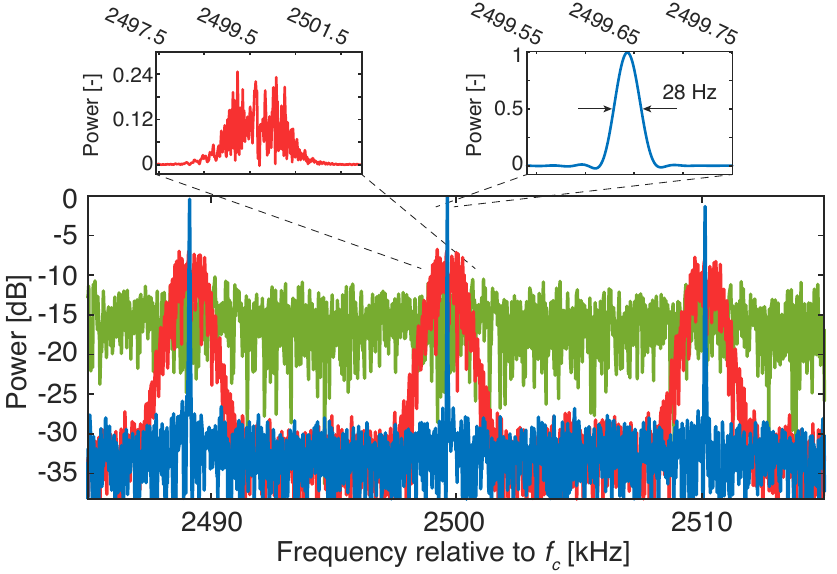}
\caption{Evolution of a small region of the beat spectrum computed from a 71-ms-long IGM stream as it goes through the different correction steps. Green: Raw spectrum. Red: Spectral shifting compensated after phase correction. Modes close to $f_c$ are nearly perfect at this stage. Blue: Spectral stretching compensated after resampling. The insets show the evolution of one mode, which finally becomes transform-limited. All vertical axes are normalized to the same value.}
\label{fig:combmodes}
\end{figure}

\begin{figure}[htbp]
\centering
\includegraphics[width=\linewidth]{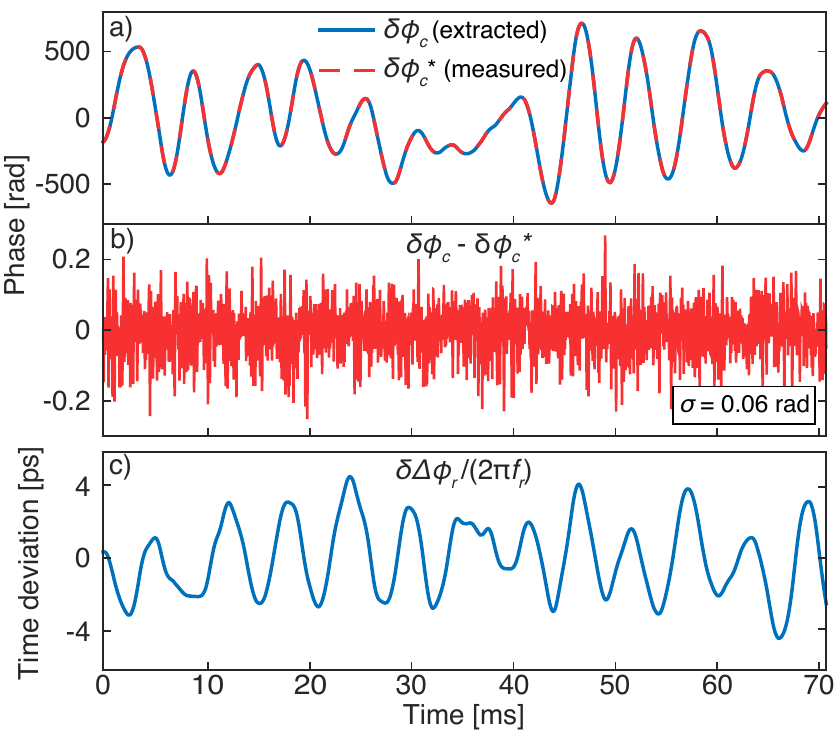}
\caption{a) Extracted phase excursions $\delta \! \phi_c(t)$ (blue) associated with the fluctuations on the RF comb's central frequency. Independent measurement of the same quantity (red), denoted as $\delta \! \phi_c^*(t)$, measured through an intermediate CW laser. b) Difference between $\delta \! \phi_c(t)$ and $\delta \! \phi_c^*(t)$. c) Extracted phase excursions $\delta \! \Delta \! \phi_r(t)$ associated with the fluctuations on the repetition rate difference. They are normalized by $2\pi f_r$ to obtain time deviations from a linear delay grid.}
\label{fig:phaseexcursions}
\end{figure}

Initially, the uncorrected spectrum is completely smeared, as shown by the green trace in Fig. \ref{fig:combmodes} computed from a 71-ms-long IGM stream. This highlights the fact that, in the original spectrum, RF modes are wider than their nominal spacing. We first compensate spectral shifting on the RF comb using a correction based on the values $\delta \! f_{c,k}$. They are used to estimate the continuous phase signal $\delta \! \phi_c(t) = 2\pi \int \delta \! f_c(t) dt$ required to perform a phase correction \cite{deschenes2010optical}. This corrects the fluctuations of the mode at $f_c$ ($n=0$), but leaves spectral stretching around that point uncompensated, as depicted by the red higher-index modes in Fig. \ref{fig:combmodes}. Then, we use the values $\tau_k$ to construct the continuous phase signal $\delta \! \Delta \! \phi_r(t) = 2\pi \int \delta \! \Delta \! f_r(t) dt$ associated with spectral stretching. This phase signal is used to resample the IGMs on a grid where the delay between pairs of optical pulses is linearly increasing (constant $\Delta \! f_r$). This yields the blue spectrum in Fig. \ref{fig:combmodes}, which shows transform-limited modes having a width determined by the von Hann window that was used to compute all aforementioned spectra ($2/(71\times10^{-3}) = 28$ Hz). The improvement between the green and blue spectra indicates that this algorithm allows accounting for fluctuations greater than the RF mode spacing. The extracted $\delta \! \phi_c(t)$ is shown in blue in Fig. \ref{fig:phaseexcursions}a) and the extracted $\delta \! \Delta \! \phi_r(t)$, normalized by $2\pi f_r$ to obtain time deviations from a linear delay grid, is shown in Fig. \ref{fig:phaseexcursions}c). A detailed explanation of the algorithm is given in Appendix A.

To verify the exactitude of the extracted signal $\delta \! \phi_c(t)$, we compare it with an independent measurement of this quantity that we refer to as $\delta \! \phi_c^*(t)$. It was obtained from the beat note between two comb modes, one from each comb, measured through an intermediate CW laser. This corresponds to the approach that is routinely taken to post-correct dual-comb IGMs \cite{deschenes2010optical}. Since the pair of optical modes that is selected by the CW laser creates a beat note at a RF frequency $f_{CW}$ different from $f_c$, the phase signal corresponding to this beat note is adjusted with the term $\delta \! \Delta \! \phi_r(t)$ scaled by $(f_c-f_{CW})/\Delta \! f_r$, the number of modes separating $f_{CW}$ from $f_c$. This yields the measured signal $\delta \! \phi_c^*(t)$ shown in red in Fig. \ref{fig:phaseexcursions}a), which should give the same information as the extracted signal $\delta \! \phi_c(t)$. The difference between $\delta \! \phi_c(t)$ and $\delta \! \phi_c^*(t)$ is given in Fig. \ref{fig:phaseexcursions}b) and shows white residuals up to $\Delta \! f_r/2$, the Nyquist frequency of the sampled fluctuations. The standard deviation of the residuals is 0.06 rad, which corresponds to around one hundredth of a cycle, or 50 attoseconds at 1550 nm.

It is important to note that the algorithm presented here can only compensate relative fluctuations that are slower than $\Delta \! f_r/2$ since they are effectively sampled by each ZPD burst. Anything above this frequency is aliased during sampling and contaminates the correction signals estimated in the 0 to $\Delta \! f_r/2$ band. Thus, a high $\Delta \! f_r$ and a low level of relative frequency noise, especially in the band above $\Delta \! f_r/2$, are desirable to achieve the best results. However, $\Delta \! f_r$ must always be smaller than $f_r^2/(2 \Delta \nu)$, where $\Delta \nu$ is the optical combs' overlap bandwidth, in order to correctly map the optical information to a single comb alias. See Appendix A for the algorithm's tolerance to relative frequency noise faster than $\Delta \! f_r/2$.


Moreover, we must emphasize that the self-correction algorithm simply permits retrieving the mutual coherence between comb modes from the IGMs themselves, which yields an equidistant, but arbitrary, frequency axis. Therefore, calibration against frequency standards or known spectral features is still required if an absolute frequency axis is needed.

\section{Spectroscopy of HCN}

We use the spectrometer to measure the transmission spectrum of the $2\nu_{3}$ band of H\textsuperscript{13}C\textsuperscript{14}N by relying solely on the self-correction algorithm presented above. We mix the two frequency combs in a 50/50 coupler and send one output through a free-space gas cell (Wavelength References HCN-13-100). The 50-mm-long cell has a nominal pressure of $100 \pm 10$ Torr and is at room temperature ($22 \pm 1^{\circ}$C). The optical arrangement is such that light does three passes in the cell. The transmitted light is sent to an amplified detector (Thorlabs PDB460C-AC) while the second coupler's output goes straight to an identical detector that provides a reference measurement (see Fig. \ref{fig:setup}). This reference measurement is especially important to calibrate the spectral modulation present on the generated combs. Both signals are simultaneously acquired with an oscilloscope (Rigol DS6104) operating at 1 GS/s.

\begin{figure*}[htpb]
\centering
\includegraphics[width=\linewidth]{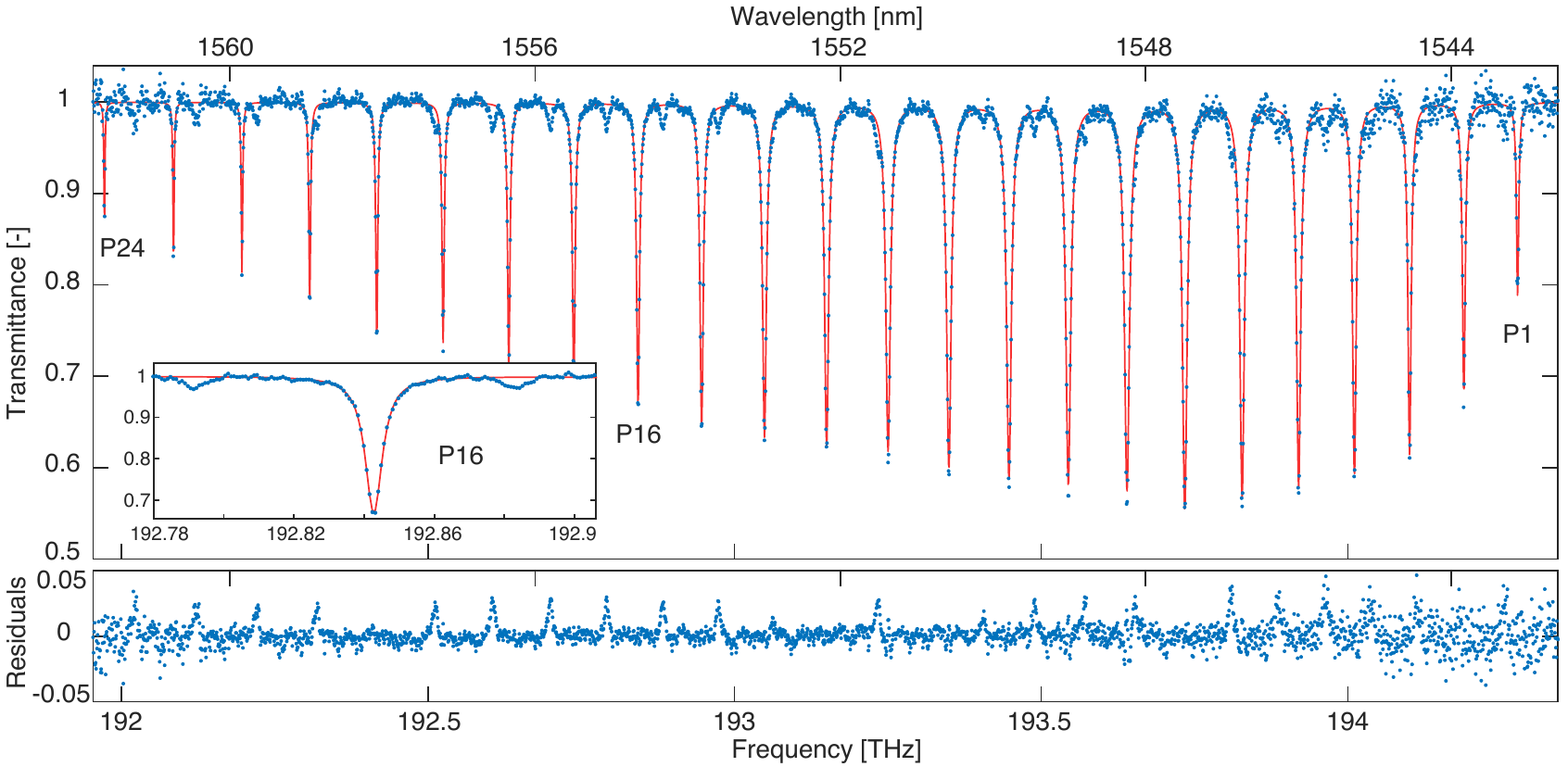}
\caption{Transmission spectrum of H\textsuperscript{13}C\textsuperscript{14}N. Each blue point corresponds to the power of a single dual-comb mode and the point spacing is $822.4$ MHz. The red curve is the result of a fit composed of 24 Voigt lines. Fit residuals show weak hot-band transitions that were not taken into account in the model.}
\label{fig:HCN}
\end{figure*}

Figure \ref{fig:HCN} shows a transmission spectrum acquired in 71 ms that covers up to 20 nm of spectral width. This allows to observe 24 absorption lines belonging to the P branch of H\textsuperscript{13}C\textsuperscript{14}N with a spectral point spacing of $f_r = 822.4$ MHz. The absolute offset of the frequency axis is retrieved by using one of the spectral features' known centre and its scale is determined by using the measured value for $f_r$, as in \cite{1612.00055}. We overlay the result of a fit composed of 24 Voigt lines for which the Gaussian width (Doppler broadening) is determined by calculation from \cite{demtroder2013laser} for a temperature of $22 ^{\circ}$C (FWHM $\sim 450$ MHz). The Lorentzian width (pressure broadening), centre and depth of each line are left as free parameters. We use the same approach as the one described in \cite{hebert2015quantitative} to fit the data and suppress the slowly varying background. The dominant structure left in the fit residuals is due to weak hot-band transitions. They represent the biggest source of systematic errors for the retrieved parameters since they often extend over lines of interest. 

\begin{table}[htpb]
 \setlength{\tabcolsep}{6pt}
\centering
\caption{Lorentzian half widths retrieved from the fit compared to reference widths calculated from \cite{swann2005line} assuming a pressure of 92.84 Torr. Uncertainties are the $2\sigma$ confidence intervals.}
\label{tab:widths}
\begin{tabular}{|c|c|c|c|}
\hline
Line & \begin{tabular}[c]{@{}c@{}}Measured width\\ {[}MHz{]}\end{tabular} & \begin{tabular}[c]{@{}c@{}}Expected width\\ {[}MHz{]}\end{tabular} & \begin{tabular}[c]{@{}c@{}}Deviation\\ {[}MHz{]}\end{tabular} \\ \hline
P1   & 2089(184)                                                               & 2042(186)                                                          & $47 \pm 370$                                   \\
P4   & 3188(118)                                                               & 3157(139)                                                          & $31 \pm 257$                                    \\
P5   & 3548(121)                                                               & 3528(93)                                                            & $20 \pm 214$                                    \\
P9   & 4170(132)                                                               & 4178(93)                                                            & $-8 \pm 225$                                     \\
P10  & 4076(133)                                                               & 4131(93)                                                           & $-55 \pm 226$                                   \\
P11  & 3946(134)                                                               & 3992(93)                                                            & $-46 \pm 227$                                   \\
P14  & 3324(134)                                                               & 3296(139)                                                          & $28 \pm 273$                                    \\
P16  & 2651(132)                                                               & 2739(139)                                                          & $-88 \pm 271$                                   \\
P17  & 2368(133)                                                               & 2414(139)                                                          & $-46 \pm 272$                                   \\
P20  & 1805(140)                                                               & 1671(232)                                                          & $134 \pm 372$                                  \\
P23  & 1035(160)                                                               & 1068(186)                                                          & $-33 \pm 346$                                   \\
P24  & 881(201)                                                                & 882(186)                                                             & $-1 \pm 387$                                     \\ \hline
\end{tabular}
\end{table}

As a final proof that the correction algorithm presented in this paper yields quality spectroscopic data, we compare Lorentzian half widths obtained from the fit to values derived from broadening coefficients reported in \cite{swann2005line}. Note that reference data is not available for all lines. We calculate reference widths from reported broadening coefficients (in MHz/Torr) using a cell pressure of 92.84 Torr, which lies within the manufacturer's tolerance. This pressure value yields minimum deviations between measured and reference widths and is in good agreement with the value of $92.5 \pm 0.8$ Torr estimated from a different experiment using the same gas cell \cite{hebert2015quantitative}. The measured and reference widths along with their deviations are gathered in Table \ref{tab:widths}. The measurement uncertainties correspond to the $2\sigma$ confidence intervals returned by the fit. The excellent agreement between the two value sets confirms the reliability of the spectrometer and of its correction algorithm. If the correction had left any significant fluctuations uncompensated, the spectrum would have appeared smeared, and the lines would have been broadened.

\section{Conclusion}

We have designed and demonstrated the use of a new kind of dual-comb spectrometer based on passively mode-locked on-chip WGLs. This platform improves the mutual stability of dual-comb systems by coupling the cavities mechanically and thermally. Combined with a new correction algorithm that allows to extract and compensate residual fluctuations, this free-running system can perform mode-resolved spectroscopy without using any external information. This self-correction approach to dual-comb interferometry can be used with any pair of combs having sufficient mutual coherence on a $1/\Delta \! f_r$ timescale.

This compact instrument and its self-correction approach represent two important steps towards the widespread adoption of dual-comb spectroscopy. The design could even be miniaturized down to a monolithic device with a SAM directly mounted on the end-face of the chip. Single- and dual-comb versions of this device could reach multi-GHz repetition rates and compete against microresonator-based frequency combs \cite{del2007optical, kippenberg2011microresonator, suh2016microresonator}. Moreover, the broad transparency of the ZBLAN chip makes the platform easily adaptable to the mid-infrared, a key enabler for useful spectroscopy applications.


\section*{Funding}
Natural Sciences and Engineering Research Council of Canada (NSERC); Fonds de Recherche du Qu\'ebec - Nature et Technologies (FRQNT).

\section*{Acknowledgments}
The authors thank Sarah Scholten from Andre Luiten's group for lending the gas cell.


\begin{thebibliography}{99}
\newcommand{\enquote}[1]{``#1''}

\bibitem{cossel2012broadband}
K.~C. Cossel, D.~N. Gresh, L.~C. Sinclair, T.~Coffey, L.~V. Skripnikov, A.~N.
  Petrov, N.~S. Mosyagin, A.~V. Titov, R.~W. Field, E.~R. Meyer \emph{et~al.},
  \enquote{Broadband velocity modulation spectroscopy of HfF+: Towards a
  measurement of the electron electric dipole moment,} Chem. Phys. Lett.
  \textbf{546}, 1--11 (2012).

\bibitem{li2008laser}
C.-H. Li, A.~J. Benedick, P.~Fendel, A.~G. Glenday, F.~X. K{\"a}rtner, D.~F.
  Phillips, D.~Sasselov, A.~Szentgyorgyi, and R.~L. Walsworth, \enquote{A laser
  frequency comb that enables radial velocity measurements with a precision of
  1 cm s-1,} Nature \textbf{452}(7187), 610--612 (2008).

\bibitem{truong2015accurate}
G.-W. Truong, J.~Anstie, E.~May, T.~Stace, and A.~Luiten, \enquote{Accurate
  lineshape spectroscopy and the boltzmann constant,} Nat. Commun. \textbf{6}
  (2015).

\bibitem{berntsson2002quantitative}
O.~Berntsson, L.-G. Danielsson, B.~Lagerholm, and S.~Folestad,
  \enquote{Quantitative in-line monitoring of powder blending by near infrared
  reflection spectroscopy,} Powder Technol. \textbf{123}(2), 185--193 (2002).

\bibitem{funke2003techniques}
H.~H. Funke, B.~L. Grissom, C.~E. McGrew, and M.~W. Raynor, \enquote{Techniques
  for the measurement of trace moisture in high-purity electronic specialty
  gases,} Rev. Sci. Instrum. \textbf{74}(9), 3909--3933 (2003).

\bibitem{maslowski2016surpassing}
P.~Maslowski, K.~F. Lee, A.~C. Johansson, A.~Khodabakhsh, G.~Kowzan,
  L.~Rutkowski, A.~A. Mills, C.~Mohr, J.~Jiang, M.~E. Fermann \emph{et~al.},
  \enquote{Surpassing the path-limited resolution of fourier-transform
  spectrometry with frequency combs,} Phys. Rev. A \textbf{93}(2), 021802
  (2016).

\bibitem{stowe2008direct}
M.~C. Stowe, M.~J. Thorpe, A.~Pe'er, J.~Ye, J.~E. Stalnaker, V.~Gerginov, and
  S.~A. Diddams, \enquote{Direct frequency comb spectroscopy,} Adv. At. Mol.
  Opt. Phy. \textbf{55}, 1--60 (2008).

\bibitem{foltynowicz2013cavity}
A.~Foltynowicz, P.~Mas{\l}owski, A.~J. Fleisher, B.~J. Bjork, and J.~Ye,
  \enquote{Cavity-enhanced optical frequency comb spectroscopy in the
  mid-infrared application to trace detection of hydrogen peroxide,} Applied
  Physics B \textbf{110}(2), 163--175 (2013).

\bibitem{marian2005direct}
A.~Marian, M.~C. Stowe, D.~Felinto, and J.~Ye, \enquote{Direct frequency comb
  measurements of absolute optical frequencies and population transfer
  dynamics,} Phys. Rev. Lett. \textbf{95}(2), 023001 (2005).

\bibitem{holzwarth2000optical}
R.~Holzwarth, T.~Udem, T.~W. H{\"a}nsch, J.~Knight, W.~Wadsworth, and P.~S.~J.
  Russell, \enquote{Optical frequency synthesizer for precision spectroscopy,}
  Phys. Rev. Lett. \textbf{85}(11), 2264 (2000).

\bibitem{coddington2016dual}
I.~Coddington, N.~Newbury, and W.~Swann, \enquote{Dual-comb spectroscopy,}
  Optica \textbf{3}(4), 414--426 (2016).

\bibitem{coddington2008coherent}
I.~Coddington, W.~C. Swann, and N.~R. Newbury, \enquote{Coherent
  multiheterodyne spectroscopy using stabilized optical frequency combs,} Phys.
  Rev. Lett. \textbf{100}(1), 013902 (2008).

\bibitem{baumann2011spectroscopy}
E.~Baumann, F.~Giorgetta, W.~Swann, A.~Zolot, I.~Coddington, and N.~Newbury,
  \enquote{Spectroscopy of the methane $\nu$ 3 band with an accurate
  midinfrared coherent dual-comb spectrometer,} Phys. Rev. A \textbf{84}(6),
  062513 (2011).

\bibitem{deschenes2010optical}
J.-D. Desch{\^e}nes, P.~Giaccari, and J.~Genest, \enquote{Optical referencing
  technique with cw lasers as intermediate oscillators for continuous full
  delay range frequency comb interferometry,} Opt. Express \textbf{18}(22),
  23358--23370 (2010).

\bibitem{roy2012continuous}
J.~Roy, J.-D. Desch{\^e}nes, S.~Potvin, and J.~Genest, \enquote{Continuous
  real-time correction and averaging for frequency comb interferometry,} Opt.
  Express \textbf{20}(20), 21932--21939 (2012).

\bibitem{ideguchi2014adaptive}
T.~Ideguchi, A.~Poisson, G.~Guelachvili, N.~Picqu{\'e}, and T.~W. H{\"a}nsch,
  \enquote{Adaptive real-time dual-comb spectroscopy,} Nat. Commun. \textbf{5}
  (2014).

\bibitem{ideguchi2016kerr}
T.~Ideguchi, T.~Nakamura, Y.~Kobayashi, and K.~Goda, \enquote{Kerr-lens
  mode-locked bidirectional dual-comb ring laser for broadband dual-comb
  spectroscopy,} Optica \textbf{3}(7), 748--753 (2016).

\bibitem{mehravar2016real}
S.~Mehravar, R.~Norwood, N.~Peyghambarian, and K.~Kieu, \enquote{Real-time
  dual-comb spectroscopy with a free-running bidirectionally mode-locked fiber
  laser,} Appl. Phys. Lett. \textbf{108}(23), 231104 (2016).

\bibitem{gong2014polarization}
Z.~Gong, X.~Zhao, G.~Hu, J.~Liu, and Z.~Zheng, \enquote{Polarization
  multiplexed, dual-frequency ultrashort pulse generation by a birefringent
  mode-locked fiber laser,} in \enquote{CLEO: Science and Innovations,}
  (Optical Society of America, 2014), pp. JTh2A--20.

\bibitem{zhao2016picometer}
X.~Zhao, G.~Hu, B.~Zhao, C.~Li, Y.~Pan, Y.~Liu, T.~Yasui, and Z.~Zheng,
  \enquote{Picometer-resolution dual-comb spectroscopy with a free-running
  fiber laser,} Opt. Express \textbf{24}(19), 21833--21845 (2016).

\bibitem{chang2015dual}
M.~Chang, H.~Liang, K.~Su, and Y.~Chen, \enquote{Dual-comb self-mode-locked
  monolithic yb: Kgw laser with orthogonal polarizations,} Opt. Express
  \textbf{23}(8), 10111--10116 (2015).

\bibitem{fork1981generation}
R.~Fork, B.~Greene, and C.~V. Shank, \enquote{Generation of optical pulses
  shorter than 0.1 psec by colliding pulse mode locking,} Appl. Phys. Lett.
  \textbf{38}(9), 671--672 (1981).

\bibitem{link2015dual}
S.~M. Link, A.~Klenner, M.~Mangold, C.~A. Zaugg, M.~Golling, B.~W. Tilma, and
  U.~Keller, \enquote{Dual-comb modelocked laser,} Opt. Express \textbf{23}(5),
  5521--5531 (2015).

\bibitem{link2016dual}
S.~M. Link, A.~Klenner, and U.~Keller, \enquote{Dual-comb modelocked lasers:
  semiconductor saturable absorber mirror decouples noise stabilization,} Opt.
  Express \textbf{24}(3), 1889--1902 (2016).

\bibitem{villares2015chip}
G.~Villares, J.~Wolf, D.~Kazakov, M.~J. S{\"u}ess, A.~Hugi, M.~Beck, and
  J.~Faist, \enquote{On-chip dual-comb based on quantum cascade laser frequency
  combs,} Appl. Phys. Lett. \textbf{107}(25), 251104 (2015).

\bibitem{rosch2016chip}
M.~R{\"o}sch, G.~Scalari, G.~Villares, L.~Bosco, M.~Beck, and J.~Faist,
  \enquote{On-chip, self-detected terahertz dual-comb source,} Appl. Phys.
  Lett. \textbf{108}(17), 171104 (2016).

\bibitem{burghoff2016computational}
D.~Burghoff, Y.~Yang, and Q.~Hu, \enquote{Computational multiheterodyne
  spectroscopy,} Science Advances \textbf{2}(11) (2016).

\bibitem{champak}
C.~Khurmi, N.~B. H\'{e}bert, W.~Q. Zhang, S.~A. V., G.~Chen, J.~Genest, T.~M.
  Monro, and D.~G. Lancaster, \enquote{Ultrafast pulse generation in a
  mode-locked erbium chip waveguide laser,} Opt. Express \textbf{24}(24),
  27177--27183 (2016).

\bibitem{schlager2003passively}
J.~B. Schlager, B.~E. Callicoatt, R.~P. Mirin, N.~A. Sanford, D.~J. Jones, and
  J.~Ye, \enquote{Passively mode-locked glass waveguide laser with 14-fs timing
  jitter,} Opt. Lett. \textbf{28}(23), 2411--2413 (2003).

\bibitem{beecher2010320}
S.~Beecher, R.~Thomson, N.~Psaila, Z.~Sun, T.~Hasan, A.~Rozhin, A.~Ferrari, and
  A.~Kar, \enquote{320 fs pulse generation from an ultrafast laser inscribed
  waveguide laser mode-locked by a nanotube saturable absorber,} Appl. Phys.
  Lett. \textbf{97}(11), 111114 (2010).

\bibitem{thoen2000erbium}
E.~Thoen, E.~Koontz, D.~Jones, F.~Kartner, E.~Ippen, and L.~Kolodziejski,
  \enquote{Erbium-ytterbium waveguide laser mode-locked with a semiconductor
  saturable absorber mirror,} IEEE Photon. Technol. Lett. \textbf{12}(2),
  149--151 (2000).

\bibitem{gross2013femtosecond}
S.~Gross, D.~G. Lancaster, H.~Ebendorff-Heidepriem, T.~M. Monro, A.~Fuerbach,
  and M.~J. Withford, \enquote{Femtosecond laser induced structural changes in
  fluorozirconate glass,} Opt. Mater. Express \textbf{3}(5), 574--583 (2013).

\bibitem{minoshima2001photonic}
K.~Minoshima, A.~M. Kowalevicz, I.~Hartl, E.~P. Ippen, and J.~G. Fujimoto,
  \enquote{Photonic device fabrication in glass by use of nonlinear materials
  processing with a femtosecond laser oscillator,} Optics Letters
  \textbf{26}(19), 1516--1518 (2001).

\bibitem{davis1996writing}
K.~M. Davis, K.~Miura, N.~Sugimoto, and K.~Hirao, \enquote{Writing waveguides
  in glass with a femtosecond laser,} Opt. Lett. \textbf{21}(21), 1729--1731
  (1996).

\bibitem{choudhury2014ultrafast}
D.~Choudhury, J.~R. Macdonald, and A.~K. Kar, \enquote{Ultrafast laser
  inscription: perspectives on future integrated applications,} Laser Photon.
  Rev. \textbf{8}(6), 827--846 (2014).

\bibitem{smart1991cw}
R.~Smart, D.~Hanna, A.~Tropper, S.~Davey, S.~Carter, and D.~Szebesta,
  \enquote{Cw room temperature upconversion lasing at blue, green and red
  wavelengths in infrared-pumped Pr3+-doped fluoride fibre,} Electron. Lett.
  \textbf{27}(14), 1307--1309 (1991).

\bibitem{palmer2013high}
G.~Palmer, S.~Gross, A.~Fuerbach, D.~G. Lancaster, and M.~J. Withford,
  \enquote{High slope efficiency and high refractive index change in
  direct-written Yb-doped waveguide lasers with depressed claddings,} Opt.
  Express \textbf{21}(14), 17413--17420 (2013).

\bibitem{lancasterer3+}
D.~G. Lancaster, Y.~Li, Y.~Duan, S.~Gross, M.~W. Withford, and T.~M. Monro,
  \enquote{Er3+ active Yb3+ Ce3+ co-doped fluorozirconate guided-wave chip
  lasers,} IEEE Photon. Technol. Lett. \textbf{28}(21), 2315--2318 (2016).

\bibitem{lancaster2015holmium}
D.~G. Lancaster, V.~J. Stevens, V.~Michaud-Belleau, S.~Gross, A.~Fuerbach, and
  T.~M. Monro, \enquote{Holmium-doped 2.1 $\mu$m waveguide chip laser with an
  output power $>$ 1 W,} Opt. Express \textbf{23}(25), 32664--32670 (2015).

\bibitem{lancaster2013efficient}
D.~G. Lancaster, S.~Gross, H.~Ebendorff-Heidepriem, M.~J. Withford, T.~M.
  Monro, and S.~D. Jackson, \enquote{Efficient 2.9 $\mu$m fluorozirconate glass
  waveguide chip laser,} Opt. Lett. \textbf{38}(14), 2588--2591 (2013).

\bibitem{nagamatsu2004influence}
K.~Nagamatsu, S.~Nagaoka, M.~Higashihata, N.~Vasa, Z.~Meng, S.~Buddhudu,
  T.~Okada, Y.~Kubota, N.~Nishimura, and T.~Teshima, \enquote{Influence of Yb3+ 
  and Ce3+ codoping on fluorescence characteristics of Er3+-doped fluoride
  glass under 980 nm excitation,} Opt. Mater. \textbf{27}(2), 337--342 (2004).

\bibitem{woodward2014probability}
P.~M. Woodward, \emph{Probability and Information Theory with Applications to
  Radar} (Permagon, 1953), 2nd ed.

\bibitem{1612.00055}
G.-W. Truong, E.~M. Waxman, K.~C. Cossel, E.~Baumann, A.~Klose, F.~R.
  Giorgetta, W.~C. Swann, N.~R. Newbury, and I.~Coddington, \enquote{Accurate
  frequency referencing for fieldable dual-comb spectroscopy,} Opt. Express
  \textbf{24}(26), 30495--30504 (2016).

\bibitem{demtroder2013laser}
W.~Demtr{\"o}der, \emph{Laser spectroscopy: basic concepts and instrumentation}
  (Springer Science \& Business Media, 2013).

\bibitem{hebert2015quantitative}
N.~B. H{\'e}bert, S.~K. Scholten, R.~T. White, J.~Genest, A.~N. Luiten, and
  J.~D. Anstie, \enquote{A quantitative mode-resolved frequency comb
  spectrometer,} Opt. Express \textbf{23}(11), 13991--14001 (2015).

\bibitem{swann2005line}
W.~C. Swann and S.~L. Gilbert, \enquote{Line centers, pressure shift, and
  pressure broadening of 1530-1560 nm hydrogen cyanide wavelength calibration
  lines,} J. Opt. Soc. Am. B \textbf{22}(8), 1749--1756 (2005).

\bibitem{del2007optical}
P.~Del'Haye, A.~Schliesser, O.~Arcizet, T.~Wilken, R.~Holzwarth, and
  T.~Kippenberg, \enquote{Optical frequency comb generation from a monolithic
  microresonator,} Nature \textbf{450}(7173), 1214--1217 (2007).

\bibitem{kippenberg2011microresonator}
T.~J. Kippenberg, R.~Holzwarth, and S.~Diddams, \enquote{Microresonator-based
  optical frequency combs,} Science \textbf{332}(6029), 555--559 (2011).

\bibitem{suh2016microresonator}
M.-G. Suh, Q.-F. Yang, K.~Y. Yang, X.~Yi, and K.~J. Vahala,
  \enquote{Microresonator soliton dual-comb spectroscopy,} Science
  \textbf{354}(6312), 600--603 (2016).


\end{thebibliography}

\section*{Appendix A: Detailed self-correction algorithm}

The algorithm aims to correct both degrees of freedom on the RF comb: its spectral spacing and its spectral offset. This is done by extracting the values $\tau_k$ and $\delta \! f_{c,k}$ for each $k^{th}$ ZPD burst using the cross-ambiguity function and by deriving the continuous phase signals $\delta \! \phi_c(t)$ and $\delta \! \Delta \! \phi_r(t)$ in order to perform a correction as the one described in \cite{deschenes2010optical}. We start by shifting the spectrum to DC with a phase ramp having the slope of the first IGM's carrier frequency. This slope is evaluated with a linear fit to the phase ramp in the first ZPD burst.  We then interpolate the values $\delta \! f_{c,k}$, which are measured at ZPD times deduced from the values $\tau_k$, in order to obtain $\delta \! f_{c}(t)$ for all times. In other words, we simply interpolate the value pairs $(\tau_k,\delta \! f_{c,k})$ using a spline. We then integrate $\delta \! f_{c}(t)$ to retrieve the associated phase signal $\delta \! \phi_{c,1}(t)$ and use it to apply a first phase correction on the IGM stream. This operation corrects most of spectral shifting and starts to reveal the comb's modal structure. Although they can be distinguished, the modes still occupy a significant fraction of the mode spacing. At this point, the spectrum's centre of mass is aligned with DC because of the spectral shift that was initially applied. The mode closest to DC is the mode $n=0$, which was initially at frequency $f_c$.

Since this first correction signal was obtained by integrating interpolated frequency data, it did not necessarily force the right set of phase values at ZPD times. Therefore, we can refine the phase correction further by extracting the residual phase excursions in the IGM stream. To do so, we cross-correlate the first ZPD burst with the rest of the IGM stream, which is safe now that most $\delta \! f_c(t)$ is compensated, and extract each burst's residual phase offset $\phi_k$. As long as the first correction was seeded with adequately sampled fluctuations, the $\phi_k$ values now have sufficiently small jumps ($\phi_k - \phi_{k-1}<\pi$) so that they can be unwrapped. However, excess noise can be found on the $\phi_k$ values in the case where the dual-comb system exhibits relative frequency noise faster than $\Delta \! f_r/2$, which cannot be accounted for during the first correction. Therefore, this algorithm can tolerate a certain amount of relative frequency noise in the band above $\Delta \! f_r/2$ as long as it translates into noise on the $\phi_k$ values that satisfies $\phi_k - \phi_{k-1}<\pi$. 
The value pairs $(\tau_k, \phi_k)$ are unwrapped and interpolated to create a second phase signal $\delta \! \phi_{c,2}(t)$, which we use for a second phase correction that fully corrects the mode $n=0$ to a transform-limited peak at DC. The sum $\delta \! \phi_{c,1}(t) + \delta \! \phi_{c,2}(t) = \delta \! \phi_c(t)$ represents the complete signal that would have been required to perform a one-off correction from the start. This step completes the correction of spectral shifting, but leaves spectral stretching uncompensated. Note that the red trace in Fig. 5 corresponds to the spectrum corrected incrementally with both $\delta \! \phi_{c,1}(t)$ and $\delta \! \phi_{c,2}(t)$ or, equivalently, directly with $\delta \! \phi_c(t)$. The blue curve in Fig. 6a) corresponds to the complete signal $\delta \! \phi_c(t)$.

Next, we define a phase vector that represents the evolution of the repetition rate difference. We set the phase to 0 at the first ZPD time and increment it by $2\pi$ at successive ZPD times. This is justified by the fact that the arrival of ZPD bursts is periodic and each burst indicates the start of a new IGM. We interpolate the value pairs $(\tau_k, 2\pi(k-1))$ for all times and remove the linear trend on the resulting signal, which yields the continuous phase fluctuations $\delta \! \Delta \! \phi_r(t)$. This data can finally be used to construct a resampling grid for the IGM stream where the delay between pairs of optical pulses is linearly increasing (constant $\Delta \! f_r$). This resampling correction compensates spectral stretching around the mode $n=0$ at DC.

\end{document}